\documentstyle[revtex]{aps}
\begin{document}
\draft
\begin{title}
Low voltage conductance in small tunnel junctions.
\end{title}
\author{Francisco Guinea\cite{csic}}
\begin{instit}
The Harrison M. Randall Laboratory of Physics, The
University of Michigan, Ann Arbor, Michigan 48109-1120
\end{instit}
\author{Masahito Ueda}
\begin{instit}
NTT Basic Research Laboratories, 3-9-11 Midori-cho, Musashino-shi,
 Tokyo 180, Japan
\end{instit}
\begin{abstract}
A discrete charge transfer in a small tunnel junction where Coulomb
interactions are important can excite electron-hole pairs
near the Fermi level.
We use a simple model to study the associated nonequilibrium properties and
found two novel effects:
(i) for junctions with electrodes of the same electronic properties,
a leakage current exists within the Coulomb gap even when the environmental
impedance is infinite;
(ii) for junctions with electrodes of different electronic properties,
the differential conductance diverges when a net interaction between
conduction electrons is attractive, and it is strongly suppressed for a net
repulsive interaction.
\end{abstract}

\narrowtext

Small tunnel junctions show unique features associated with a discrete charge
transfer across the energy barrier\cite{Likh}.
These systems are being studied in
detail, and have striking implications for other fields of
physics \cite{Zeit}.

In this Letter, we focus on tunneling processes occurring at energies
below the elementary charging energy $E_C = e^2 /2 C$, $e$ and $C$ being the
electronic charge and  the junction capacitance.
This is the energy range within which small tunnel junctions show their most
interesting properties, such as Coulomb blockade.
Our main results in this regime are:

i) Junctions whose electrodes
have the same electronic properties, show a leakage
current within the Coulomb gap even when the environmental
impedance is infinite. This leakage current smoothes the current-voltage (I-V)
characteristics and reduces the Coulomb offset extrapolated back from the
high-voltage part of the I-V curve.

ii) In junctions with electrodes of different electronic properties, the I-V
characteristics show rectifying features within the Coulomb gap, and under
certain circumstances, conductivity diverges at zero voltage, exhibiting a
superconducting-like behavior.

While most discussions are restricted to single junctions, this Letter also
presents extensions to multijunction arrays.

A distinguishing feature of our analysis is that it takes account of the fact
that electron tunneling in small junctions strongly perturbs the electronic
states close to the Fermi level.
The conventional theory assumes that the junction is in a new equilibrium
state immediately after a tunneling event.
This implies that the electronic wavefunctions which
describe the electrons near the Fermi level are not affected by the
change in the charge state of the junction, and only a rigid shift
in the energy levels must be considered.
The validity of this assumption, in fact, depends on the inhomogeneities in
the potential generated after the charge build-up, over scales comparable to
the mean free path of the electrons.
When the inhomogeneities cannot be neglected, the dependence of the electronic
wavefunctions on the charge state of the junction must be included
in the calculations. The importance of this effect increases as the
electrodes become smaller.

If the change in the electronic wavefunctions is not negligible, each
charge transfer is followed by sizable shake-up processes \cite{Ueda}.
The Nozi\`{e}res-Dominicis problem describes, in a similar way, the
ejection of a core electron which changes the charge state of an atom
within a metal \cite{ND}.
A substantial shakeup of the conduction electrons follows,
leading to singularities in the photoemission and X-ray
absorption spectra. As in the junction, the system experiences a
fast change --the modification of the charge state-- and a resulting
slow relaxation.
The main alteration in the junction case is that the transferred electron
that ^^ triggers' a time-dependent perturbation is itself part of the Fermi
liquid.
This difference, however, does not qualitatively change the above picture
because the final state interaction is caused, not by the tunneling
electron itself, but the change in the {\it collective}
coordinate which describes the charge state.

For more precision, we separate the tunneling event into two processes
having very different time scales:
(i) an electron near the Fermi level in one electrode tunnels to the other,
giving rise to a charge build-up, and
(ii) low-energy electron-hole pairs are excited in response to this
quick change in the potential.
This idea can be implemented mathematically by separating the degrees of
freedom of the junction into two classes:
a collective one, which lumps together all ^^ fast' variables,
and the dynamics of the electrons at the Fermi level, which can be described
in terms of independent electron-hole pairs. The dynamics of the
fast variable are associated with changes in the charge state
of the junction. We define this collective variable as $Q$ and
 its conjugate variable as $\phi$
such that $[ \phi , Q ] = i e$, where $e$ is the electronic charge.
The wavefunction of the junction can be written as a
superposition of states specified by the variable $\phi$ and the
occupancy of the levels around the Fermi energy:
$ | \phi , \{ n_k \} >$.
Charge relaxation occurs over energy scales
comparable to the plasmon energy, $\sim 1$eV, while
we use the quasiparticle description for processes at energies, or voltages,
comparable to the charging energy of the system, $E_C \sim 1 - 10$meV,
so that our distinction is justified.

 Including all these effects, the junction
can be described by the Hamiltonian \cite{Ued2},
\begin{eqnarray}
{\cal H} &= &{\cal H}_C + {\cal H}_L + {\cal H}_R +
{\cal H}_T + {\cal H}_V \nonumber \\
{\cal H}_C &= &{{{(Q - Q_x )}^2} \over {2 C}} \nonumber \\
{\cal H}_{R,L} &= & \sum_k \epsilon_{k,R,L} c^+_{k,R,L}
c_{k,R,L} \nonumber \\
{\cal H}_T &= & \hat{t} e^{i \phi} \sum_{k,k'} c^{+}_{k,R}
c_{k',L} + h. c. \nonumber \\
{\cal H}_V &= & v_R ( Q - Q_x ) \sum_{k,k'} c^{+}_{k,R} c_{k',R}
\nonumber \\ &
& - v_L ( Q - Q_x ) \sum_{k,k'} c^{+}_{k,L}  c_{k',L}
\end{eqnarray}
Here $Q_x$ stands for any offset charge which may exist at equilibrium.
The differences between Hamiltonian (1) and more conventional
treatments are the separation of slow and fast degrees of
freedom, and  the introduction of the shakeup processes through
the last term \cite{Ueda}.
Note that the tunneling hamiltonian ${\cal H}_T$ is defined such that
modification in the charge state is accompanied by a
change in the distribution of quasiparticles at the Fermi level.

We can eliminate the slow degrees of freedom,
leaving only the charge. The influence of the
electrons on the evolution of the charge
can be expressed in terms of
electronic propagators \cite{Guin}.
The inclusion of the shake-up processes implies that the propagators
have to be computed for the {\it nonequilibrium} ground state of charged
electrodes.
We will use only  their asymptotic behavior at long times
\cite{Ueda}:
\begin{equation}
G^{R,L} ( t ) = - i {{e^{- i E_F t}}\over{t}}
( E_C t )^{g_{R,L}}
\end{equation}
In equation (2), the effects of the potential on the electrons
at the Fermi level, due to $v_R$ and $v_L$, are described by the
phase shifts induced at the Fermi
level, $\delta_{R,L}$, and
$g_{R,L} = 2 \delta_{R,L} / \pi -( \delta_{R,L} / \pi )^2$.

The contribution that has the anomalous exponent arises from the shake-up.
An effective coupling between tunneling processes separated by time $t$
is described by the product  $G^R ( t )  \times G^L ( t )$,
which depends on a single parameter,
 $\bar{g} = - g_R - g_L$.
This approach, which is based on the definition of an appropriate tunneling
density of states that includes the effects of the interaction, is rather
general, and can be applied in many contexts \cite{Fish}.
\vskip 0.5cm
We now consider the case of a single junction with electrodes of the same
 electronic properties.
After a tunneling event, the potentials acting on the states at the
Fermi level are equal in magnitude but opposite in sign, and so are the phase
 shifts, i.e., $\delta_R = - \delta_L \equiv \delta$.
Hence, we find that $\bar{g} = 2 (\delta/\pi)^2 > 0$.
In the small coupling limit, $t / E_C \ll 1$, the current
operator, $e \partial ( N_R - N_L ) / \partial t$, is
\begin{equation}
\hat{j} = i e \hat{t} e^{i \phi} \sum_{k,k'} c^{+}_{k,R} c_{k',L} +
{\rm h.c.}
\end{equation}
Using Kubo's formula, the conductivity at finite frequencies
and zero temperature is given by
\begin{equation}
\sigma ( \omega ) =
{{ \sum | < 0 | \hat{j} | \omega > |^2}\over{\omega}}
\end{equation}
where the summation is taken over all excited states of energy $\omega$.

The numerator in (4) is the Fourier transform of a correlation function
of the type $< \hat{j} ( t ) \hat{j} ( t' ) >$.
We now express this correlation
function as a series expansion in powers of ${\cal H}_T$. Each insertion
of this operator generates a kink in the time evolution of the charge.
A schematic representation of a given term in this series is shown
in Fig. 1.

We can use the standard analogy with a  one dimensional system
with logarithmic interactions \cite{PWA}.
The contribution of the path shown in Fig. 1 then reduces to the
interaction effects between two charge dipoles at separation $t$.
It decays over a long time like $t^{(-4-\bar{g})}$. Taking the
Fourier transform, and inserting it in (4), we obtain
\begin{eqnarray}
\sigma ( \omega ) & \sim {{e^2}\over{h}}
t^4 N_R ( E_F )^2 N_L ( E_F )^2 \left( {{\omega}\over{E_C}}
\right)^{2 + \bar{g} } \nonumber \\
& \sim {{e^2}\over{h}} \left( {{R_Q}\over{R_T}} \right)^2
\left( {{\omega}\over{E_C}} \right)^{2 + \bar{g} }.
\end{eqnarray}
where $N_R ( E_F )$ and $N_L ( E_F )$ denote the respective
density of states at the Fermi level in the right and left electrodes.
This formula is valid at low frequencies, $\hbar \omega \ll E_C$.

The existence of a finite conductance at frequencies
within the Coulomb gap also implies a finite differential resistance
in the same region. Equation (5) is a measure of the probability
that the junction absorbs energy in quanta given by $\hbar \omega$.
Thus, at low voltages, we expect $\hbox{I} \propto \hbox{V}^{3 + \bar{g}}$,
as shown in Fig. 2.

Our results imply that a single normal tunnel junction is always dissipative.
The decay of the ^^ Coulomb blockade state' cannot be significantly reduced
by lowering the temperature or increasing the resistance in the external
circuit.
We stress that, in an experiment designed to measure the I-V characteristics,
this current within the Coulomb gap does not violate the conservation law
of energy.
To determine the I-V characteristics, the junction has to be attached to a
reservoir in which dissipation can take place.
Otherwise, even in the absence of Coulomb blockade, electrons at
the electrodes cannot be described by Fermi-Dirac distributions.
When the circuit is attached to a macroscopic battery, processes
which do not change the electrochemical potential difference $V$ across
the junction are not forbidden because, as in the M\"{o}ssbauer effect,
the battery instantaneously compensates the work $eV$.
This situation is similar to what we have encountered in the low-impedance
environment. In this case, spontaneous fluctuations in the junction charge
due to the electromagnetic environment induces tunneling within the Coulomb
gap; this process, however, does not violate the conservation law of energy
because the battery again compensates the energy necessary for tunneling
\cite{Devoret}.
In the high-impedance case, however, the second-order process in tunneling
is inhibited, and therefore our fourth-order process comes to play a major
role.

The same result can be expressed in terms of the fluctuation-dissipation
theorem. To make the chemical potentials at the two electrodes unequal,
the system has to be brought out of equilibrium. The
perturbation required to do so has an intrinsic time scale, given
by $\hbar /$eV (a voltage can be induced by an ac magnetic field,
for instance). Processes which have longer relaxation times
play no role in the steady state reached by the system.
That the absence of these processes give rise to a voltage- or
temperature-dependent renormalization of the junction capacitance
has been discussed extensively.
By the same analysis, these processes can also contribute to the imaginary
part of the response function, leading to the dissipation we have shown above.

The existence of a finite subgap conductance will
also reduce the Coulomb offset extrapolated back from the high-voltage part of
 the I-V curve.
Most of the current will come from tunneling processes which involve energies
 between $E_C$ and the applied
voltage. They will give rise to a linear dependence of
I on V. A current due to the small subgap conductance also
contributes, reducing the Coulomb offset.
Both the smoothing of the I-V characteristic and the reduction in the Coulomb
offset seem to be in agreement with experimental observations
\cite{Dels,Kuzm,Clel}. Finally, dissipation will be accompanied
by heating effects \cite{Garc}.

The analysis presented here can be readily extended to
many junction arrays. The frequency (or voltage) dependence
of the conductivity is not changed because
it is simply related to the interaction between two
charge dipoles in a manner similar to that depicted in Fig. 1.
The prefactor, on the other hand, has to be replaced by
$( h / e^2 R )^N$. This term increases exponentially
with the number of junctions for high resistance
junction arrays. It will compete with other processes already discussed
in the literature, like macroscopic quantum tunneling of charge
\cite{Aver,Geer,Hann}. Physically,
the latter
implies two coherent tunneling events at two different junctions,
and is drastically different from the shakeup we discuss here.
It seems an interesting question to separate these two
sources of subgap conductance. From the requirement of coherence,
we conclude that macroscopic tunneling
of charge should be very sensitive to changes in the conductance of
a single junction. Shakeup processes, on the other hand, should
always be present, independently of the ratio between the
different conductances.
\vskip 0.5cm
We next consider junctions with electrodes of different electronic properties
 for which $\delta_R \ne \delta_L$. We concentrate
on the high conductance limit, $\alpha = h / e^2 R \gg 1$. In this case,
it is convenient to start with the effective action:
\begin{eqnarray}
S_{eff} &= \sum_{\pm} \alpha_{\pm} \tau_c^{\pm ( g_R + g_L )}
\nonumber \\
&\int^{\beta}_0 d \tau \int^{\beta}_0 d \tau ' {{1 - e^{\pm i
[ \phi ( \tau ) - \phi ( \tau ') ]}}\over{(\tau - \tau '
)^{[ 2 \pm ( g_R + g_L ) ]}}} + 2\pi i {{ Q_x} \over {e}}
\end{eqnarray}
where the parameter $\tau_c$ is the cutoff which is initially set equal to the
 charging energy.
Since the two possible electron jumps between the electrodes are
described by different propagators, they are associated with the two sign
 choices in the action. The first term is complex,
indicating an instability of the ^^ neutral' junction. A real
partition function is recovered by assuming a temperature-dependent
offset charge $Q_x$ at the electrodes. We have
 defined two different conductances, $\alpha_{\pm}$. At high
voltages or temperatures, we have $\alpha_+ = \alpha_- =
\alpha_0$, where  $\alpha_0 = h / ( e^2 R )$.

The high frequency modes in the phase can be
integrated out in (6), leading to the  scaling equations
\cite{Kost}
\begin{eqnarray}
{{\partial \alpha_{\pm}}\over{\partial \ln (eV) }}
&= &\pm \alpha_{\pm}
 + {2\over{\pi^2}} \nonumber \\
{{\partial Q_x}\over{\partial \ln (eV) }} &= &- \alpha_{-}
( g_R + g_L)
\end{eqnarray}
Here, the second equation is needed to keep the partition function real
at all scales. The first equation is obtained by scaling
separately the contributions from the two types of hopping
processes.

These equations have a nontrivial fixed point at $\alpha_{{\rm crit}} =
2 / ( \pi^2 | g_R + g_L | )$. For initial conductances
above this value, the conductances within the Coulomb gap scale
towards infinity as the scale of frequencies (or voltages) is
reduced. Below this threshold, we recover the scaling equations for
symmetric tunnel junctions already known \cite{Guin}.
Integrating (7), we obtain

\[ \begin{array}{rcrrr}
\alpha_{\pm} ( V ) & = & \alpha_0 \left( {{eV}\over{E_C}}
\right)^{\pm (g_R + g_L)} & \qquad \alpha_0 \gg \alpha_{{\rm crit}}   \\
\alpha_{\pm} ( V ) & = & \alpha_0 - {2\over{\pi^2}} \log
\left( {{E_C}\over{eV}} \right)
& \qquad \alpha_0 \ll \alpha_{{\rm crit}}  & \qquad (8)
\end{array} \]

The first of these equations leads to asymmetric I-V characteristics.
The differential resistance at zero vanishes for
 the positive sign and diverges for the negative
sign. The qualitative features of the I-V characteristics in the two
cases are shown in Fig. 3.

The scaling equations (7) are valid in the limit of
high conductances. This condition is always satisfied when the
frequency (or voltage) is scaled towards lower values in the high-conductance
side of the I-V characteristics. This effective ^^ normal-state
superconductivity' is a novel effect which merits further theoretical and
 experimental study.

The range of parameters required for its observation can be inferred from the
previous analysis. The difference in phase shifts between the two electrodes
must be large. This implies that they must be very different, and that the
potentials induced by their respective charge distributions must be local in
 space.
A negative phase shift indicates that an additional charge pulls states below
the Fermi level, while a positive shift has the opposite effect. One of
the  electrodes should therefore show a net repulsion between charges, while
the other should have a net attraction. Similarly,
it can be shown that a divergent
conductivity arises from tunneling between
two attractive Luttinger liquids \cite{Fish}.
Finally, the
normal state conductance, $\alpha_0$, has to be large, meaning good
contact between the electrodes, or many channels
through which electrons can tunnel. The latter possibility, however, will
tend to reduce the phase shifts by spreading the potential of
a unit charge over the different channels. We therefore think that this
effect would be better observed in small tunnel junctions with
good connections.

We appreciate interesting discussions with E. Ben-Jacob. F.G.
 also acknowledges financial support from CITyT (grant MAT 88-0211).

\figure{Path integral formulation of the calculation
of the $< \hat{j} ( t' )
\hat{j} (t) >$ correlation function (see text).}

\figure{I-V characteristics of a normal, symmetric tunnel junction.}

\figure{I-V characteristics of an asymmetric tunnel junction, where
$V_c= E_C/e \exp (-\pi^2(\alpha_0-1)/2)$ is the renormalized offset voltage.
Solid  line: $\alpha_0 \ll \alpha_{{\rm crit}}$.
Broken line: $\alpha_0 \gg \alpha_{{\rm crit}}$.}

\begin{references}
\bibitem[*]{csic}Permanent Address: Instituto de Ciencia de
Materiales. Consejo Superior de Investigaciones
Cient{\'i}ficas. Universidad Aut\'onoma de Madrid. E-28049 Madrid. Spain.

\bibitem{Likh}D. V. Averin and K. K. Likharev, in
{\it Mesoscopic Phenomena in Solids}, B. L. Altshuler,
P. A. Lee and R. A. Webb eds. (North-Holland, Amsterdam, 1991), p.173.

\bibitem{Zeit}Special Topics issue of Zeitschrift f\"{u}r Physik B on
{\it Single Charge Tunneling},
{\bf 85} December 1991 (Proceedings of the NATO ASI, Les Houches, France,
 5-15 March, 1991).

\bibitem{Ueda}M. Ueda and S. Kurihara, NTT preprint; in {\it Macroscopic
Quantum
 Phenomena}, eds. T. D. Clark, H. Prance, R. J. Prance and T. P. Spiller (World
 Scientific, Singapore, 1991), p.143.

\bibitem{Guin}F. Guinea and G. Sch\"on, Europhys. Lett. {\bf 1}, 585 (1986).

\bibitem{ND}P. Nozi\`eres and C. T. De Dominicis, Phys. Rev. {\bf 178}, 1097
(19
   69).

\bibitem{Ued2}M. Ueda and F. Guinea, in ref. 2, p. 413.

\bibitem{Fish}C. L. Kane and M. P. A. Fisher,
Phys. Rev. Lett. {\bf68} 1220 (1992).

\bibitem{PWA}P. W. Anderson and G. Yuval, J. Phys. C {\bf 4}, 607 (1971).

\bibitem{Devoret} M. H. Devoret, D. Esteve, H. Grabert, G.-L. Ingold,
H. Pothier, and C. Urbina, Phys.~Rev.~Lett. {\bf 64}, 1824 (1990);
S. M. Girvin, L. I. Glazman, M. Jonson, D. R. Penn, and M. D. Stiles,
Phys.~Rev.~Lett. {\bf 64}, 3183 (1990).

\bibitem{Dels}D. B. Haviland, L. S. Kuzmin, P. Delsing, K. K. likharev
and T. Claeson in ref. 2, p. 339.

\bibitem{Kuzm}L. S. Kuzmin, Yu. V. Nazarov, D. B. Haviland, P. Delsing and T.
Claeson, Phys. Rev. Lett. {\bf 67} 1161 (1991).

\bibitem{Clel} A. N. Cleland, J. M. Schmidt and John Clarke,
Phys. Rev. B {\bf 45} 2950 (1992).

\bibitem{Garc}N. Garc{\'\i}a and F. Guinea, Phys. Rev. B, in press.

\bibitem{Aver}D. V. Averin and A. A. Odintsov, Phys. Lett. A {\bf 140}, 251
(198
   9).
Y. V. Nazarov, in ref. 2.

\bibitem{Geer}L. J. Geerligs, D. V. Averin and J. E. Mooij, Phys. Rev. Lett.
{\bf 65}, 3037 (1990).

\bibitem{Hann}A. E. Hanna, M. T. Tuominen and M. Tinkham,
Phys. Rev. Lett. {\bf 68} 3228 (1992).

\bibitem{Kost}J. M. Kosterlitz, Phys. Rev. Lett. {\bf 37}, 1577 (1976).




\end{references}
\end{document}